\documentclass[conference]{article}
\usepackage[hmargin=.75in,vmargin=1in]{geometry}
\usepackage[american]{babel}
\usepackage[T1]{fontenc}
\usepackage{times}
\usepackage{textcomp}
\usepackage{epsfig,graphicx}
\usepackage{amsfonts,amsmath,amssymb}
\usepackage{fixltx2e} 

\title{\bf  An eco-friendly Ecash with recycled banknotes}           

\author{
{\bfseries Amadou Moctar Kane}\\
KSecurity BP 47136 Dakar Senegal\\
amadou1@gmail.com
}

\begin{document}

\maketitle                        

\begin {abstract}

By comparing cryptocurrencies with other existing payment methods, including banknotes and bank cards, it is clear that the use of Bitcoin and its competitors (Ethereum, \dots) is almost insignificant in world trade.  We may also note that these cryptocurrencies have become tools of speculation, which is the antithesis of their primary purpose. 
 Based essentially on the security of electronic signatures, the Ecash introduced here will put the users back to the center of the game and exclude miners and their enormous waste of power energy. 
Thus, the purpose of this paper is to show that even a piece of paper can be recycled into a secure Ecash, while remaining environmentally friendly. Hence, we create here a cryptocurrency that would use a slight modification of the current banknotes to set up anonymous electronic transactions. 
By trading with banknotes, we mechanically transfer ownership of the paper money from one owner to another, hence, in this scheme, we introduce the notion of ownership transfer. It implies that at each transaction the elements allowing to authenticate the Ecash does not change, while the ownership certificate will change since the Ecash is transfer towards someone else. 

\end {abstract}

\medskip
\section{ Introduction}
Gold is valuable because it is a rare ore, the day when each of us could make tons of gold, this metal would be worthless.  
This is the problem of most of existing cryptocurrencies, since there exists currently a proliferation of Ecash \cite{faure}.  Why should we let our neighbors or colleagues make tons of gold which they would sell us at exorbitant prices?  

Bitcoin and its sisters were originally intended to put an end to the banks' hold on currencies, but today they have become a kind of electronic gold that is not suited to our current transactions. 

Similarly, the mining of some cryptocurrencies is energy consuming, in January 2018, for Bitcoin, it was estimated between 21 and 52 billion KWh per year \cite{bitcoin}.

However, it should be noted, that Bitcoin is highly appreciated by the public, although it faces number of attacks \cite{conti}. 

\subsection{Related work}

The design of the electronic cash has been very beneficial for cryptography since the creation of the first Ecash in 1982 by David Chaum \cite{chaum}, this field of research has allowed the introduction of several tools whose most emblematic have been the blind signature and the blockchains. 

In his first scheme, Chaum had created a system that allowed the user to withdraw an electronic coin from a bank and spend the coin with a merchant without his bank identifying it. During the transaction, the merchant can verify the authenticity of the electronic coins (in some protocols the merchant does not need to interact with the bank before accepting a coin from the user).

In 2008, the introduction of Bitcoin and its blockchain technology by Satoshi \cite{nakamoto} created a huge enthusiasm in the cryptocurrencies area, it also created a large amount of new e-cash such as Ethereum \cite{buterin}. 
In parallel with the blockchain technology, completely different schemes will be introduced with the implementation of the compact e-cash \cite{camenisch} which permits the withdrawal of $2^l$ coins in a short time. 

It has also been introduced the malleable signature \cite{chase}, where the authors create the Ecash by delegating the signing authority of a TTP which would allow anonymous transaction without the involvement of the TTP. 

Later, Tewari et al. \cite{tewari} used the blind signature protocol to anonymise users and blockchain and proof of work algorithms to prevent double spending and to verify the authenticity of transaction. unfortunately this interesting scheme will need miners  (rewarded by the bank), blockchain and proof of work which could create the inherent problems of Bitcoin such as power energy waste.

More recently, Lipton et al. \cite{lipton} have proposed the use of blockchain technology to track the relevant transaction parameters, reducing the opportunity for parties to be dishonest. While payments are still direct between users as in Chaum's proposal. Unfortunately, the Ecash tracking ledger reintroduce in that scheme is not a real time protection against a double-spending and could lead to a massive fraud before any audit. 

As said before, even if we have now plenty of cryptocurrencies such as Bitcoin, we do not have Ecash for our daily transaction online due to the fact that cryptocurrencies fall far short of being true currencies \cite{rogoff} and can not replace the very useful banknotes. Instead of replacing the banknotes, these cryptocurrencies have become highly speculative tools which could transform the live of its users in a nightmare within a short time\cite{caginalp}, hence they can not be useful as Ecash. 

Credit cards are designed in a way that allows governments, and private corporation such as facebook \cite{facebook}, to store and to trace the user's privacy to a degree never seen before. Hence, these can not be useful for our privacy.

Only banknotes are widely accepted and allow us to have an interesting level of security and privacy, thus protecting us from data predators. It is the reason why we present here, a secure electronic cash, based on a slight modification of the classical banknotes. 

\subsection{Our contribution}
In \cite{dice}, an identifiable banknote (RFID / Code) connects to a digital security system to verify the banknote's validity.  A key feature of this technology is the ability to devaluate banknotes that may be stolen from a user or illegally circulating.  However, our approach in this paper is different if we add the possibility for users to verify the authenticity of a banknote or an Ecash, we also protect the users against any intervention of the authorities during the payment process. 

Since the first scheme of Ecash of David Chaum, most of the schemes we have seen worked hard to protect the Ecoins, for example against the double spending, while the approach we take here is completely different since the Ecash that we propose can be copied an infinity number of times (including by an adversary), however the certificate of ownership that all users of the Ecash possesses can only be modified by its legitimate owner. 

Signatures by their native functions (non-repudiation, authenticity, data integrity \dots) will protect users against fraud. 

In this paper, we propose an Ecash which can be in an electronically version or in a paper version.  We will use a database to store the certificates of ownership containing in particular the public key of the owner of the Ecash.  

The database will evolve progressively, because at each transaction, the ownership certificate concerned change, it is replaced by the new one (we remove the old file after each transaction). 

The currency has been defined by Aristotle \cite{aristote, wikipedia} by three functions: unit of account, store of value and medium of exchange. Even if this definition has evolved today, it contains some functions which are still interesting.

Hence, in order give our Ecash as a store of value function, the issuer may be required to have sufficient funds to repay part of the losses of users of the electronic cash in the case of technological failures attributable to it, or a sharp drop of the currency. 

In comparison with banknotes, we are aiming to give to the user of the Ecash at least the same level of security and privacy.

\subsection{Organization of the Paper}

This paper is organized as follows, in Section 2 we will describe our new scheme, in section 3 we present an example of our scheme and before the conclusion we will present our security and efficiency analysis.

\section{The electronic cash scheme}

\subsection{Notation}

\begin{itemize}
	\item We denote by $Sign(X:Y)$, the Signature obtained from a private key encryption such that $X$ is the plaintext, and $Y$ the encryption key.
	\item A message may have several components and message components will be separated by ||. Thus $Sign({N||A}: SK_{U_{1}})$ denotes that the message signed by the key $SK_{U_{1}}$ is composed by the two parts $N$ and $A$.
	
	\item In this article we will have the issuer of the Ecash (CashCreator) instead of the bank, thus, the issuer can be a commercial bank, a natural or a legal person, a group of people (they will use a group signature to sign the certificates), a central bank or any other entity presenting the guarantees necessary for the issue of a currency. 
		
	\item As mentioned above the issuer of the currency will sometimes be noted $CC$, the users of the Ecash would be noted $U$, since we can have several, it could be numbered , for example $U_1$ would be the first user.  
	
	\item We denote by $H$, the hash function used in this scheme and we suppose that it is a collision resistant hash function. It is called A collision resistance hash function due to the fact that it is computationally hard to find two inputs $a$ and $b$ which are different and which produce the same output $H(a)=H(b)$.
	
\item  $SK_{U_{1}}$ will be $U_1$'s private key and $PK_{U_1}$ will be his public key. $SK_{CC}$ will be $CC$'s private key and $PK_{CC}$ its public key.

\item Each Ecash has a number which is unique, we denote this number by $N$ the, $AM$ is the amount of the Ecoin and $P$ is the picture of the e-cash (it can be the current banknote, or a selfie of the user).

\end{itemize}
Any e-cash is composed by three parts:
\begin{itemize}
	\item Part A: the certificate of ownership. 

It certifies that the possessor of the Ecash is the owner legitimate. It is subject to changes during the transaction.
	\item Part B: It is composed by a picture that could look like the banknote we currently use (with a unique number and an amount), this part is frozen and is not subject to any change once the Ecash is issued . 
		
	\item Part C: A certificate of authenticity 

 It allows the CashCreator to sign part $ B$ with his private key, this part is also frozen and is not subject to any change once the Ecash is issued. 
\end{itemize}
 
\subsection{Design}

We suppose that all communications between the protagonists are secure and we also assume that the CashCreator public key is available for all users. 

The solution proposed in this paper can be divided in three parts.
\subsubsection{The CashCreator's preparation}
	
\begin{enumerate}
	\item $CC$ creates the part $B$ of the Ecash by choosing an image ($P$), adding an amount ($AM$) and a unique number ($N$).
	
	\item $CC$ creates the part $C$ by signing with his private key the hash of part $B$. The hash is obtained with $H$.
\end{enumerate}

\subsubsection{The user's preparation}

\begin{enumerate}
	\item $U_1$ generates a pair of public and private key and sends to the CashCreator his public key.
	\item $ CC $ creates the part $A$ of the Ecash by writing its unique number, the owner's public key and by signing the whole (part $A$) with his private key. 
	\item $CC$ sends the Ecash consisting of the three parts $( A||B||C)$ to $U_1$.
	\item $U_1$ signs the part $A$ and sends it in broadcast to all users. 
	\item All users of the network $ U_2, U_3, \dots U_n $ check if the signature of $ CC $ is valid, i.e. if the signer has the right to sign it and if the signature is correct.  If the tests are conclusive, they enter Part $A$ in their database. 
	
\end{enumerate}

\subsubsection{The transfer (the spending)}

\begin{enumerate}
	\item	In order to accept the payment, the merchant denominated by $ U_2 $ sends his public key $ U_1 $.
	\item The user $U_1$ creates a new part $A$ of his Ecash by writing the number of the Ecoin $N$ and the public key of $U_2$ (instead of its own public key) and signs that part $A$ with his private key. 
	\item The merchant verifies if the last owner of this Ecash was the possessor of $U_1$'s public key in its database. And with that public key he will verify if the signature of $U_1$ is correct. if it is correct, he will sign the part A and send it to all the users of the networks.

\end{enumerate}

\textbf{Summary: }

(1) 

$U_1 \, sends \xrightarrow{\{My \, public \, key \, is \, PK_{U_{1}}\}} CashCreator$

\medskip

(2) 

The CashCreator creates the different parts of the Ecash:

$Part A = \{PK_{U_{1}} \, is \, the \, owner \, of \, the \, Ecash \, N || sign(PK_{U_{1}} \, is \, the \, owner \, of \, the \, cash \, N: \, SK_{CC}) \}$

$Part B = \{Picture || Number \, of \, the \, Ecash \, $N$ || Amount\, of \, the \, Ecash \, $AM$\}$

$Part C = \{Sign(part B : SK_{CC})\}$

\medskip

$The \, CashCreator \, sends \xrightarrow{\{Part A || Part B || Part C\}} U_1$

\medskip

(3) 

$U_1$ verifies if the two signatures are correct ($Sign(part B: SK_{CC})$ \, and \, $sign(PK_{U_{1}} \,is \, the \, owner \, of \, the \, Ecash \, N: SK_{CC})$). 

If it is correct,  he signs Part A and send by broadcast to all the users the signed part A which is 

$sign(PK_{U_{1}} \, is \, the \, owner \, of \, the \, Ecash \, N || sign(PK_{U_{1}} \, is \, the \, owner \, of \, the \, Ecash \, N: SK_{CC}): SK_{U_{1}})$.

\medskip

$U_{1} \, sends \, by \, broadcast \xrightarrow{Sign(part A: SK_{U_{1}})} U_{2} \dots U_{n}$

\medskip 

(4) 

The users will verify the validity of the signer and the signature ($Sign(part A: SK_{CC})$), if correct, they write part $A$ of the Ecoin number $N$ in their database, the transaction is accepted.

\medskip 

5) 

Whenever the user $U_1$ would like to spend the Ecash he just needs to send the copie of the Ecoin to another user ($U_2$) and transfer the ownership to $U_2$ by creating a new $part A$.

\medskip 
 
(5.1) $Part \,A = \{PK_{U_{2}} \, is \, the \, owner \, of \, the \, Ecash \, N || \,  sign(PK_{U_{2}} \,is \, the \, owner \, of \, the \, Ecash \, N: SK_{U_{1}}) \}$
 
 \medskip
 
(5.2) $U_1$ sends the Ecoin (with the modified part $A$) to $U_2$.

 \medskip

$U_1 \, sends \xrightarrow{\{Part A || Part B || Part C\}} U_2$

\medskip

$U_2$ will verify the validity of the two signatures ($Sign(part B: SK_{CC})$ and $sign(PK_{U_{2}} \, is \, the \, owner \, of \, the \, Ecash \, N: SK_{U_{1}})$), if $U_1$ is not the owner of the Ecash number $N$ in $U_2$'s database, or if the signature is incorrect then the transaction will be rejected.
\medskip

(5.3) If the verifications are conclusive $U_2$ will sign $Part A$ and send it  broadcast to all the users. 

\medskip

$U_{2} \, sends \, by \, broadcast \xrightarrow{\{Sign(part A: SK_{U_{2}})\}} U_{1} U_{3} \dots U_{n} \dots$

\medskip

Users will verify that the former owner ($U_1$) has signed the ownership transfer $Sign(part A: SK_{U_{1}}))$ and depending on the security level, they may also verify that the new owner $U_2$ has accepted the ownership transfer $Sign(part A: SK_{U_{2}}))$

\subsubsection{Remarks}
\begin{enumerate}
	\item The database hold by users contains only the parts $ A $ of the Ecash due to resources limitation (storage space , computing time, \dots). In order to reduce the size of the database, each part $A$ of an Ecash replaces the previous one, however it will depend on the user to decide whether or not he wants to keep the previous part $A$. 
	 	
	\item Normally, each user has his own database or a file to store the ownership certificates (the parts $ A $) which are constituted of three small lines (Ecash's number and public key of the owner , signature of the owner, signature of the new owner). 

However, it is possible (and even recommended) for community members to make their database available in read mode to other users. Users can also pool their resources (CPU, storage space , \dots) to share a single database. 

	\item For an Ecash that has just been issued, the first signatory of Part $A$ is necessarily the CashCreator and Part $A$ will be inserted into the database by all users. 
	
	\item There is no link between the user's identity and his public key, he should hide his IP when sending the messages (public key, broadcast, \dots). 
	
	\item Only the owner of the private key corresponding to the public key appearing on the certificate of ownership can proceed with a transfer of ownership.

	\item We assume that the database can not be manipulated or modified by a third party, in other words each user is responsible for the security of its database and its updates. 
	
\end{enumerate}

\section{Example}

Let's suppose that Harpagon has decided to give all his money (which was buried) to wikiinfo, that Ecash is composed by banknotes. 
\begin{enumerate}
	\item We suppose that Harpagon had received a banknote along with the certificate of authenticity and the certificate of ownership (certificates can be cash receipts).

	\item As soon as he had received the banknotes, He had verified with the public key of the bank that the electronic signature of the bank affixed on the ownership certificate and that on the certificate of authenticity are valid. 
	
	\item He replaces his public key that was written on part A by the public key of wikiinfo , and signs part A with his private key.

		\item He takes a selfie of the Ecash (the banknote and the two certificates) and sends the selfie to wikiinfo via the TOR network \cite{Dingledine}. 
		
		\item wikiinfo checks in its database that the public key 01 6E F4 DB FC 9C 63 96 01 EC 84 was the holder of the rights on the banknote number 12345. 
		
	\item By using part $C$, wikiinfo verifies with the public key of the bank, that part $B$ (picture) has been signed by the bank. 
	
	\item wikiinfo signs part $A$ with its private key and sends it in broadcast mode to all users of the network.

	\item Organized by trusted network, each users group pools their resources to check the validity of the transaction.  Thus, they check that the ticket belonged to the public key 01 6E F4 DB FC 9C 63 96 01 EC 84 and that the transfer of ownership property was signed by that key.  If the verification is correct they enter the transaction in their database, otherwise they reject it. 

\end{enumerate}

\textbf{Remarks}

\begin{enumerate}

	\item Using the selfie must obey standards (pixel, camera angle, etc.) to have the same result in terms of signature. 
	
	\item If Aragon managed to steal Harpagon's Ecash, he would not be able to use it since he needs Harpagon's private key to use the stolen ticket.  On the other hand, Harpagon could use the stolen ticket if he kept a copy or a selfie of the stolen ticket.	
	\item Harpagon can also be allowed to choose the picture (part $B$) of the Ecash, by sending his selfie to the CashCreator along his public key. Hence in a case of loss or theft he could re-create the Ecash without the help of the CashCreator.
\end{enumerate}

\section {Security and efficiency analysis}

\subsection{Security analysis}

\textbf{Definition }

In this scheme, the Unforgeability states that valid coins can only be issued by the CashCreator.
Secondly, the anonymity ensures that a user stays anonymous even if the complete system conspires against him. 
Finally, the exculpability is obtained when a malicious CashCreator should not be able to conspire with malicious users to frame an honest user for double-spending.

We consider in this paper that an Ecash is secure if it has the four following properties:	
\begin{enumerate}
	\item The Ecash is unforgeable.
	\item The Ecash user is anonymous.
	\item The Ecash scheme has an exculpability property.
	\item The Ecash is protected against the double-spending.
\end{enumerate}
In the following, we will prove that our scheme respects these four properties under the assumption that the signature used in our scheme is unforgeable. We suppose that we have three potential attackers in this scheme, the attack can be tried by the user, by the CashCreator, or by someone else who is completely outside the scheme named Aragon. 
\subsubsection{Unforgeability}

As said before, the first transaction must be signed by the CashCreator, hence, any new Ecash which is not signed by the CashCreator will be rejected. Therefore, under the assumption that the CashCreator signature is unforgeable and the hash function used is collusion-resistant, The unforgeability is effective for someone who is outside the Ecash scheme (Aragon).
The unforgeability is also effective for the user due to the reasons cited previously.
In the case where the CashCreator is not entitled to create an unlimited number of banknotes, it is sufficient to count the number of notes present in the   database to see if it has exceeded the allowed number. 

Let's assume that the bank tries to swindle the user.

The bank has an accomplice who is a user, the bank issues a banknote to the user $ U_1 $ and issues the same note to his accomplice $ U_2 $. Both Ecash have the same unique number, the same amount and the same certificate of authenticity however the certificate of ownership is different since on the first Ecash it would put the public key of $ U_1 $ and on the second note (fraudulent) it would put the public key of its accomplice. 

This attack will fail, since $U_1$'s database is updated and secured. 
Another protection will be to verify in other users' database before finalizing his transaction. If $ U_1 $ signs the ownership certificate and sends it to broadcast to all users, then he will verify if that transaction does appear in other users' public database, if not, he will assume that his own database has been attacked and refuse to finalize the transaction.
If after the acceptation of $U_1$'s transaction by users, $U_2$ (bank accomplice) signs and sends the part $A$ of his Ecash, then the transaction of the accomplice would be rejected since the CashCreator is no longer the owner of the Ecash but $U_1 $ so the transfer of ownership should be signed by $ U_1 $ and not by the CashCreator. 

\subsubsection{Anonymity}

The bank can register the unique number of the Ecash and the public key of the first owner in order to trace the use of the electronic cash. 

It is important to note that the public key given to the bank is not a certificate with the name of the user, it is rather a completely anonymous public key that can belong to anyone.  However if the bank manage to identify the owner of a public key, it is enough to the owner to hide his IP and to create a new key using any tools such as OpenSSL \cite{openssl} and to  transfer the ownership of the Ecash to the new public key.

Let's suppose that after an investigation, the government has been able to seize a number of Ecash (banknotes \& certificates), it will never be able to use it, to destroy it or to know if these notes have passed into other hands or if they are still held by the person with whom they have been found (we just need a picture/selfie of the note and the associated private key to spend the Ecash).  Thus, in the absence of the private key, the anonymity of the user is preserved even if the ticket is found in his possession. Similarly, the government can not destroy Ecash that are online and which it does not know the real owner. 

\subsubsection{The Double-spending}

Let's suppose that $ U_1 $ has transferred the ownership of the Ecash number 12345 to $ U_2$ and $ U_3 $, the first one $ U_2 $ will sign and send by broadcast the ownership certificate to all users. When $U_3$ verifies his updated database he will find that the ownership has change, hence $U_1$ can not spend this Ecash the transaction will be rejected implying that a double spending should not be possible. 
 
Can the attacker blocks the network of $ U_3$ to prevent him to know that the Ecash has already been transferred. 

This attack is not in the scope of this paper, since we always assume that the user has protected his tools (e-wallet, computer, smartphone, database, networks \dots) against any attackers, hence the attacker should not be able to prevent $U_3$ from receiving the modification of the ownership.

As said before, $U_3$ should also be able to consult some database put online by other users to detect the attempt of double spending of $U_1$.

\subsubsection{exculpability}

The double spending is not feasible for the user, then the exculpability is effective in this scheme.

\subsection{Efficiency Analysis}
The evaluation method used here is also used in \cite{kane} in order to perform our critical discussion.  We prove in this section that our scheme satisfies the specific features of usability, and we compare our scheme with the current banknote. 

\subsubsection{P2P Transferability:} 
The first advantage of this Ecash is its transferability, since this scheme is essentially based on the transfer of ownership. 

\subsubsection{Interoperability:} 
This system allows the use of current banknotes in electronic money and vice versa, because the user has just to prove the ownership of the Ecash (banknote) and to transfer that ownership to someone else through an electronic signature 
\subsubsection{Applicability, Efficiency and Cost:} 
This transfer of ownership costs is very low if users pool their resources, for example if one million users pool their resources they will need to use a hash function once and to verify a signature once, for each transaction. They also will need storage space (database) and a connection for the network members (for example internet).  Hence, a small smartphone could verify and make transactions. 
The time need for computing the signature and the hash is low. We can add that the storage need is not important because it consists on storing approximately 3 short sentences for each transaction. 
\subsubsection{Ease of use:} 
stages of preparation, and transfer are very short. It is totally different from the cumbersome Bitcoin' s mining. 
\subsubsection{Scalability:} 
This system is scalable, even the people's selfies or current banknotes (from all countries) can be used in this scheme, since it is enough to issue them an electronic signature and to sign the ownership electronically in order to transform them into an Ecash. However the storage space and the signature system is limited by the of users capabilities (if they pool their resources, it will be easier).
\subsubsection{Off-line usage:}
 The system works in one-line mode due to the fact that we need to verify the ownership certificate of the user. 

\subsubsection{Mobility:} 
The important things are the signatures of the CashCreator and the users; These latter can be written in a piece of paper or kept electronically. These two tools have the property of mobility.

\subsubsection{Comparison with banknote}
Our aim was to design an Ecash which give to its user at least, the same level of security and privacy when we compare it with the banknote, hence in this subsection we will compare the two.
Part of the following properties has been defined in \cite{wang}, The traditional cash (banknote) has the following properties.
\begin{enumerate}
	\item Cash must be created such that is hard to forge it.
	\item Cash must be physically transported from one place to another place.
	\item Cash must be stored safely.
	\item Banknotes can be easily destroyed.
	\item people can steal it from other people.
	\item Cash can not be trace easily.
	\item Cash allows money laundering.
	\item Cash cannot be used for payments over the phone or the Internet.
	\item Counterfeiting banknotes can be given by the bank or a bank employee to the user. 

\end{enumerate}

The Ecash presented in this paper will fulfill the following properties.
\begin{enumerate}
	\item The signature done with the private key is hard to forge.
	
	\item The second item is solved in our scheme since the Ecash can be physically transported (if it is in the paper format) or electronically transported, however we will need a database connected with the rest of the users.
	\item The private keys must be stored safely, but the Ecash itself doesn't need to be store safely, the user just need to have at least a copy or a selfie of his Ecash.
	\item This Ecash is not easy to destroy due to the fact that the attacker needs also to destroy all the copies of that Ecash.
	\item It is more difficult to steal this Ecash, because the attacker will need a picture of the Ecash (banknote) and the private key of the owner.
	\item The Ecash can not be trace easily due to the anonymity of the public key (if use online with an anonymous network).
	\item This Ecash can allow money laundering.
	\item This Ecash can be used for payments over the phone or the Internet, it can also be used in a physical store, where the merchant can interact with others users in order to verify the ownership and to send the new certificate of ownership.
	\item Companies working in funds transfer would be useless.     
	
	\item Counterfeiting banknotes (Ecash) can not be given by the bank or a bank employee to the user, because this latter can verify live with the public key of the bank the validity of the banknote.
\end{enumerate}

\section{Conclusion}
In this article, we created an electronic cash  that can interact with our current banknotes and that can be exchanged many times.        

We have also  eliminated  in our scheme, miners who have nothing to do in a scheme of sale or transfer of funds which is a private action concerning users. 

By  eliminating  blockchains and proof of work, we have proved that the  creation  of an electronic money certainly does not require an indecent waste of energy , since with a piece of paper and a Smartphone one can set up an efficient Ecash. 

With the  advent  of the low cost cryptoprocessors of small size and large capacity \cite{dice, ibm}, it could in the future, be easier to work without the broadcast mode and the database of all the ownership certificates. By adding in banknotes the cryptoprocessor  necessary  to the  security  and  storage  of the ownership and authenticity certificates, we would have notes  issued  by the  CashCreator, that could be  used  in  e-commerce and in the shops that are on the street.

\end{document}